\documentclass[aps,pre,twocolumn,superscriptaddress,showkeys]{revtex4}
\usepackage{graphics}
\usepackage[pdftex]{graphicx}
\usepackage{times}
\usepackage{multirow}
\usepackage{pifont}
\usepackage{color}
\usepackage{float}
\usepackage[pdftex]{hyperref}
\hypersetup{colorlinks=false,pdfborder=0}
\usepackage{color}
\begin{document}

\title{Robust clustering of languages across Wikipedia growth}

\author{Kristina Ban}
\affiliation{Faculty of Information Studies, Ljubljanska cesta 31a, SI-8000 Novo mesto, Slovenia}

\author{Matja{\v z} Perc}
\thanks{Electronic address: \href{mailto:matjaz.perc@uni-mb.si}{\textcolor{blue}{matjaz.perc@uni-mb.si}}}
\affiliation{Faculty of Natural Sciences and Mathematics, University of Maribor, Koro{\v s}ka cesta 160, SI-2000 Maribor, Slovenia}
\affiliation{CAMTP -- Center for Applied Mathematics and Theoretical Physics, University of Maribor, Mladinska 3, SI-2000 Maribor, Slovenia}

\author{Zoran Levnaji\'c}
\affiliation{Faculty of Information Studies, Ljubljanska cesta 31a, SI-8000 Novo mesto, Slovenia}
\affiliation{Department of Knowledge Technologies, Jo\v zef Stefan Institute, Jamova 39, SI-1000 Ljubljana, Slovenia}

\begin{abstract}
Wikipedia is the largest existing knowledge repository that is growing on a genuine crowdsourcing support. While the English Wikipedia is the most extensive and the most researched one with over five million articles, comparatively little is known about the behavior and growth of the remaining 283 smaller Wikipedias, the smallest of which, Afar, has only one article. Here we use a subset of this data, consisting of 14962 different articles, each of which exists in 26 different languages, from Arabic to Ukrainian. We study the growth of Wikipedias in these languages over a time span of 15 years. We show that, while an average article follows a random path from one language to another, there exist six well-defined clusters of Wikipedias that share common growth patterns. The make-up of these clusters is remarkably robust against the method used for their determination, as we verify via four different clustering methods. Interestingly, the identified Wikipedia clusters have little correlation with language families and groups. Rather, the growth of Wikipedia across different languages is governed by different factors, ranging from similarities in culture to information literacy.
\end{abstract}


\maketitle

\section{Introduction}
The fact that we are able to carry the knowledge from previous generations forward gives us evolutionary advantages that no other species on the planet can compete with.
In fact, widespread cooperation among unrelated individuals on one hand, and our language on the other hand, are the two defining features that distinguish us most prominently from other species \cite{miller_81, axelrod_84, nowak_11, hrdy_11}. For millennia, we have been upholding a cumulative culture, which lead to an exponential increase in our cultural output \cite{lehman_47}. Our ability to pass on knowledge from generation to generation relies on the evolution of language \cite{nowak_n02, abrams_n03, lieberman_n07, loreto_np07, sole_jrsi10} via a set of grammatical rules that allow infinitely many comprehensible formulations \cite{chomsky_65, lightfoot_99}. In times of unprecedented technological progress and scientific breakthroughs, the amount of information to carry forward is staggering, and it requires information sharing, worldwide collaboration, algorithmic prowess of search engines, as well as selfless efforts of countless volunteers to maintain, categorize, and help navigate what we know. Wikipedia \cite{wikipedia} is surely the most famous example of what can come of such efforts.

Thankfully, much of what we know has been digitized \cite{evans_s11, michel_s11}, and the deluge of digital data, along with recent advances in the theory and modeling of social systems and networks \cite{albert_rmp01, lazer_s09, castellano_rmp09, estrada2012, kivela_jcn14, boccaletti_pr14, percm, wang2016statistical}, enables quantitative explorations of our culture that were unimaginable even a decade ago. From enhanced disease surveillance \cite{althouse2015enhancing}, human mobility patterns \cite{gonzales_n08, palchykov2014inferring}, and the spreading of misinformation \cite{bessi2015trend, del2016spreading}, to the universality in voting behavior \cite{chatterjee2013u} and emotional blogging \cite{mitrovic2010networks, mitrovic2011quantitative} to name just some examples, there are virtually no limits to innovative data-driven research that lifts the veil on how we share information, how and with whom we communicate, to where we travel, and essentially on how we live our lives.

Wikipedia \cite{wikipedia} has itself been subject to much research scrutiny, both in terms of the accuracy of content \cite{giles2005internet, chesney2006empirical,zha_pnas16}, which proved to be better than that of traditional encyclopedias, as well as in terms of intellectual interchanges during its history \cite{yun2016intellectual}, the evolution of its norm network \cite{heaberlin2016evolution}, the dynamics of conflicts and edit wars \cite{yasseri2012dynamics}, dynamics of general growth \cite{voss2005,suh2009}, circadian patterns of editorial activity \cite{yasseri2012circadian}, language complexity \cite{yasseri2012practical}, and even in terms of its converge of academics \cite{samoilenko2014distorted}. Indeed, the open access policy of Wikipedia, along with a useful API and few limits on data accessibility has made it one of the most researched data repositories in history.

Importantly, while initially research on Wikipedia has been focused predominantly on the English language, recently the focus has been shifting also towards other languages \cite{eomplos15, yasseri13, yu2014pantheon, iniguez2014modeling}, and in particular to the cross-cultural dimension of the database. In \cite{kim2016understanding}, the authors have studied editing behaviors in multilingual Wikipedia, focusing on the engagement, interests, and language proficiency in the primary and secondary languages of the editors. Research revealed that the English edition of Wikipedia displays different dynamics from the Spanish and German editions, and that it plays the broker role in bringing together content written by multilinguals from many language editions. The study also concluded that language remains a formidable hurdle to the spread of content. Similarly, in \cite{samoilenko2016linguistic}, cultural borders on Wikipedia through multilingual co-editing activity have been studied, showing that the domination of the English language disappears in the network of co-editing similarities, and that instead local connections come to the forefront. An approach has also been proposed there that allows the extraction of significant cultural borders based on the editing activity of Wikipedia users. Most recently, the early adhesion of structural inequality in the formation of Wikipedia has been studied as well \cite{yun2016early}.

\begin{table*}
\label{table26lang}
\begin{tabular}{c | c | c}
\hline\hline\textbf{Family}  &\textbf{Group}   &\textbf{Languages}    \\\hline\hline
 \multirow{7}{*}{\textbf{Indo-European}}
	  & \multirow{2}{*}{Germanic \ding{75}} & English (en), German (de), Danish (da),\\
	  &                                     & Swedish (sv), Norwegian (no), Dutch (nl) \\\cline{2-3}
	  & \multirow{2}{*}{Italic \ding{92}}   & Italian (it), Portuguese (pt), Spanish (es),\\
	  &                                     & French (fr), Romanian (ro)  \\\cline{2-3}
	  & \multirow{2}{*} {Slavic \ding{58}}  & Russian (ru), Polish (pl), Czech (cs), \\
	  &                                     & Ukrainian (uk), Bulgarian (bg), Serbian (sr)  \\\cline{2-3}
	  & Indo-Iranian \ding{108}             &  Persian (fa)  \\ \hline
	\textbf{Uralic}       & Finno-Ugric \ding{95} & Finnish (fi), Hungarian (hu) \\\hline
	\textbf{Altaid}        & Turkic \ding{90}      & Turkish (tr) \\\hline
	\textbf{Afro-Asiatic}  & Semitic \ding{217}    & Arabic (ar) \\\hline
	\textbf{Sino-Tibetian} & Chinese \ding{169}    & Chinese (zh) \\\hline
	\textbf{Korean}        &  \ding{161}           & Korean (ko) \\\hline
	\textbf{Austro-Asiatic}&  \ding{162}           & Indonesian (id) \\\hline
	\textbf{Japanese}      &  \ding{164}           & Japanese (ja) \\\hline\hline
\end{tabular}
\caption{Languages used in our study (each corresponding to one Wikipedia). All 26 languages used in our study (rightmost column) organized into language families and groups (first two columns) according to the standard reference~\cite{voegelin_77}. Italian, German and Russian for example, are all Indo-European languages, but belong to different language groups. For an easier referral a symbol is introduced in Table~\ref{tablelangclust} for each language group. Abbreviations in brackets are used later in figures and conform to the ISO 639 standard.}
\end{table*}

Here we use a relatively small subset of Wikipedia, in particular 14962 different articles, but each of which jointly exist in 26 different languages. This gives us the opportunity to study growth patterns of collaborative knowledge across time and across different languages. Essentially, we seek to explore how, given an article that exists in many Wikipedias, this article gets ``translated'' from language to language. In particular, does an average article appear in various Wikipedias following a prescribed sequence of languages or not? Can in this regard Wikipedias be clustered into language groups with shared growth properties? If yes, can these properties be understood in terms of language families, or in terms of cultural and geographical proximity, or perhaps in terms of information literacy and policy towards IT education? While we do not arrive at conclusive answers for all these questions, we do show that although an average article follows a random path from one language to another, there are nevertheless robust clusters of languages that share very similar growth rates and statistical properties of the articles' dates of birth, as well as striking similarities in the average time delays between the same articles appearing in two different languages. The languages within the identified clusters have little correlation with language families and groups, making a precise statement with regards to what exactly underpins each cluster difficult to provide.

The continuation of this paper is organized as follows. In Section 2 we first present the data that we use for our research, while in Section 3 we present the main results. We conclude in Section 4 with a discussion and an outlook into the future.

\begin{figure*}
\centering{\includegraphics[width = 16cm]{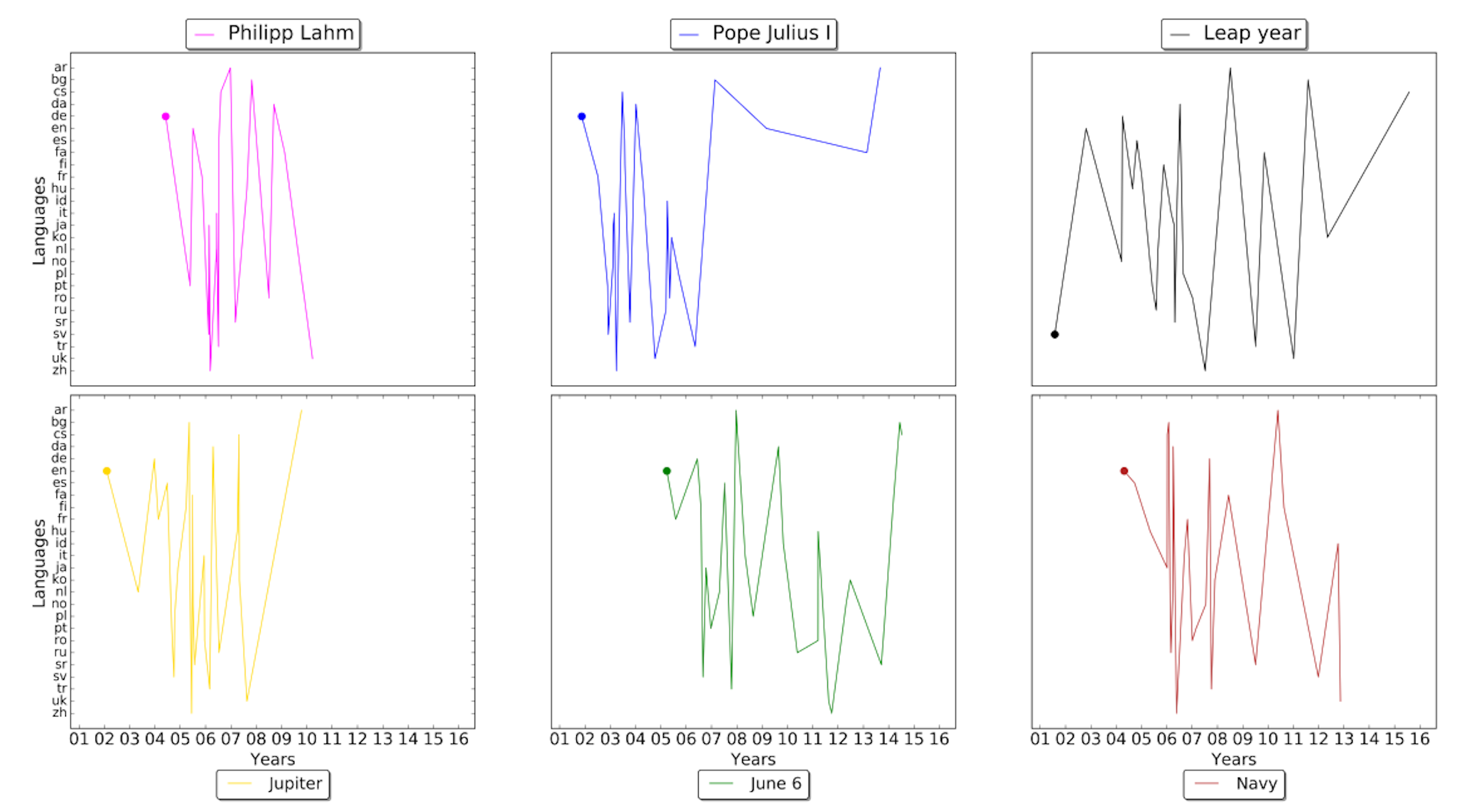}}
\caption{ An average article from Wikipedia follows a random path from the language in which first appears to all other 25 languages that our study encompasses. Shown are the trajectories of six randomly selected articles out of the whole database consisting of 14962 articles. Each trajectory connects the DOBs in all 26 Wikipedias chronologically. The DOB for the first language is marked with a circle for each article. Vertical ordering of languages is alphabetical. Time (denoted by the last two digits of each year) is displayed horizontally. The color in the trajectories is introduced solely for clarity. The Wikipedia topics of articles are shown in the legend. Presented results are representative in that randomly selecting another subset of articles would give qualitatively the same plot.}
\label{random}
\end{figure*}

\section{Data}
Non-English Wikipedias grow by both translating from other Wikipedias (typically English one) and writing articles anew. Regardless of which mechanism is predominant, what we wish to study in this paper is the spreading dynamics of articles across different language editions of Wikipedia. While we expect that the first language for the vast majority of articles is English, what we seek to elucidate is when and in which sequence those articles appear in other Wikipedias, and if there are any stable patterns in this process.

For this study we first need a suitable dataset that best allows the study of growth patterns across Wikipedias. Our first idea was to look at the articles that jointly exist in the largest number of Wikipedias. For `jointly exits' we mean that articles can be found on Wikipedia as different language versions of the same article.  We realized this identification by relying on English Wikipedia, so we first identify the English article with the corresponding articles in all other languages and then identify them all as the set of articles in different languages corresponding to each other. However, the topical diversity among such articles turned out to be very narrow (they are mostly articles for largest world countries: article for Russia is at present the most translated Wikipedia article). Also, the actual set of languages in which these articles jointly exist turns out to be rather small, which is why we deemed this dataset not suitable for our study.

Next we looked at the major Wikipedias starting with the English one, and tried to identify an ensemble of articles that jointly exist in all of them. We want to have each considered article present in each considered Wikipedia. But more Wikipedias we considered (even only large ones), smaller was the number of articles that jointly exist in all of them. This is expected, since each addition of another Wikipedia reduces the number of articles that jointly exist in all considered Wikipedias. Finally, optimizing between the two we decided to find the biggest ensemble of articles that jointly exist in the biggest set of major Wikipedias. The best situation was found for 26 Wikipedias (languages) for which there were 14962 articles that jointly exist in all of them, despite the fact that each selection leaves some important Wikipedias out. We verified that the selected ensemble of articles is topically very diverse, covering almost all domains of knowledge. The chosen 26 languages are reported in the Table~\ref{table26lang} along with their language families and groups marked by symbols. For simplicity, in this Table we also introduce an abbreviation for each language (ISO 639 standard) that we will use later.

We create the dataset by storing the date of birth (DOB) for each article in each Wikipedia. DOB we define as the date on which that article first appeared in that Wikipedia, even if it appeared as a stub (unfinished draft). We however removed the articles that never went beyond the stub stage, since they do not convey much information about Wikipedia growth. We disregard the information about further growth of these articles (as long as they were not left as stubs). Therefore, our dataset for each among 26 Wikipedia is composed of DOBs for 14962 articles. Alternatively, it can be seen as an ensemble of 14962 sequences of DOBs that correspond to 14962 articles, each sequence containing 26 DOBs corresponding to 26 languages. The articles' DOB range from the year 2001 to the year 2016. Thus formulated dataset allows for the planned study to be conducted by examining the spreading dynamics of articles from 14962 sequences of DOBs. The data was extracted and stored automatically using web crawler designed in programming language Python that was running on a parallel computer. The data was obtained in August 2016.

In the following Section we present our results, which are obtained using standard statistical analysis methods, in particular standard hierarchical clustering algorithm~\cite{hastie_09}. The clustering algorithm was implemented using programming language Python, with Euclidean distance playing the role of similarity measure (defined differently for each data analysis approach).

\section{Results}
We begin by showing in Fig.~\ref{random} trajectories over time of six randomly selected articles from our database, as they appear in language after language. The time of appearance of an articles in its first language (first DOB), which is the start of each trajectory, is marked with a circle.

It is impossible to discern a clear sequence in which the languages follow each other in succession. Articles are most likely to first appear in languages such as English or German, but apart from this rather expected observation, it is impossible to predict in which language a particular article will appear next. Also, the evolution of articles starts and finishes at different times. Thus, while there may be predictable statistical regularities in the average growth patterns (as we show later), the path an individual article takes from one language to another is random.

\begin{table}
\begin{tabular}{c | c | c | c}
\hline\hline\textbf{Growth}      & \textbf{DOB} & \textbf{Delays}& \textbf{MDS} \\\hline\hline
 English \ding{75}             & English \ding{75}            & English \ding{75}     & English \ding{75}        \\\hline
 German \ding{75}              & German \ding{75}             & German \ding{75}      & German \ding{75}      \\\hline
 Italian \ding{92}             & Italian \ding{92}            & Italian \ding{92}     & Italian \ding{92}      \\
 Finish \ding{95}              & Finish \ding{95}             & Finish \ding{95}      & Finish \ding{95}   \\
 Portuguese \ding{92}          & Portuguese \ding{92}         & Portuguese \ding{92}  & Portuguese \ding{92}          \\
 Russian \ding{58}             & Russian \ding{58}            & Russian \ding{58}     & Russian \ding{58}     \\
 Norwegian \ding{75}           & Norwegian \ding{75}          & Norwegian \ding{75}   & Norwegian \ding{75} \\
 Chinese \ding{169}            & Bulgarian \ding{58}          & Chinese \ding{169}    & Chinese \ding{169}\\
 Danish \ding{75}              & Serbian \ding{58}            & Danish \ding{75}      & Danish \ding{75}\\\hline

 Polish \ding{58}              & Polish \ding{58}             & Polish \ding{58}      & Polish \ding{58}    \\
 Dutch \ding{75}               & Dutch \ding{75}              & Dutch \ding{75}       & Dutch \ding{75}\\
 Spanish \ding{92}             & Spanish \ding{92}            & Spanish \ding{92}     & Spanish \ding{92}   \\
 Japanese \ding{164},          & Japanese \ding{164}          & Japanese \ding{164}    & Japanese \ding{164}\\
 French \ding{92}              & French \ding{92}             & French \ding{92}       & French \ding{92}\\
 Swedish \ding{75}             & Swedish \ding{75}            & Swedish \ding{75}      & Swedish \ding{75}\\
                               & Danish \ding{75}             & & \\
                               & Chinese \ding{169}           & & \\\hline
 Indonesian \ding{162}         & Indonesian \ding{162}        & Indonesian \ding{162} & Indonesian \ding{162}\\
 Turkish \ding{90}             & Turkish\ding{90}             & Turkish\ding{90}      & Turkish\ding{90}\\
 Hungarian \ding{95}           & Hungarian \ding{95}          & Hungarian \ding{95}   & Hungarian \ding{95}\\
 Korean \ding{161}             & Korean \ding{161}            & Korean \ding{161}     & Korean \ding{161}\\
 Ukrainian \ding{58}           & Ukrainian \ding{58}          & Ukrainian \ding{58}   & Ukrainian \ding{58}\\
 Czech \ding{58}               & Czech \ding{58}              & Czech \ding{58}       & Czech \ding{58} \\
 Arabic \ding{217}             & Arabic \ding{217}            & Arabic \ding{217}     & Arabic \ding{217} \\
 Romanian \ding{92}            & Romanian \ding{92}           & Romanian \ding{92}    & Romanian \ding{92} \\
 Bulgarian \ding{58}           &                              & Bulgarian \ding{58}   & Bulgarian \ding{58}\\
 Serbian \ding{58}             &                              & Serbian \ding{58}     & Serbian \ding{58}\\\hline
 Persian \ding{108}            & Persian \ding{108}           & Persian \ding{108}    & Persian \ding{108}\\\hline\hline
\end{tabular}
\caption{Clusters of  languages determined via four different methods. First column (Growth): clustering from the individual growth rates over the years (cf. Fig.~\ref{growth}); Second column (DOB): clustering from the averaged differences of DOBs (cf. Fig.~\ref{birth}); Third column (Delays): clustering from the statistics of time delays between first DOB and other DOBs (cf. Fig.~\ref{delay}); Fourth column (MDS): clustering from multi-dimensional scaling of distances between Wikipedia pairs (cf. Fig.~\ref{mds}). Different clusters are separated by horizontal lines. It can be observed that all 26 languages fall into 6 different clusters, but the make-up of the clusters changes only slightly depending on the method used. Symbols indicate language groups, as introduced in Table~\ref{table26lang}. Languages within a given cluster (where more than one) are obviously not correlated with a particular language group.}
\label{tablelangclust}
\end{table}

\begin{figure*}
\centering{\includegraphics[width = 15cm]{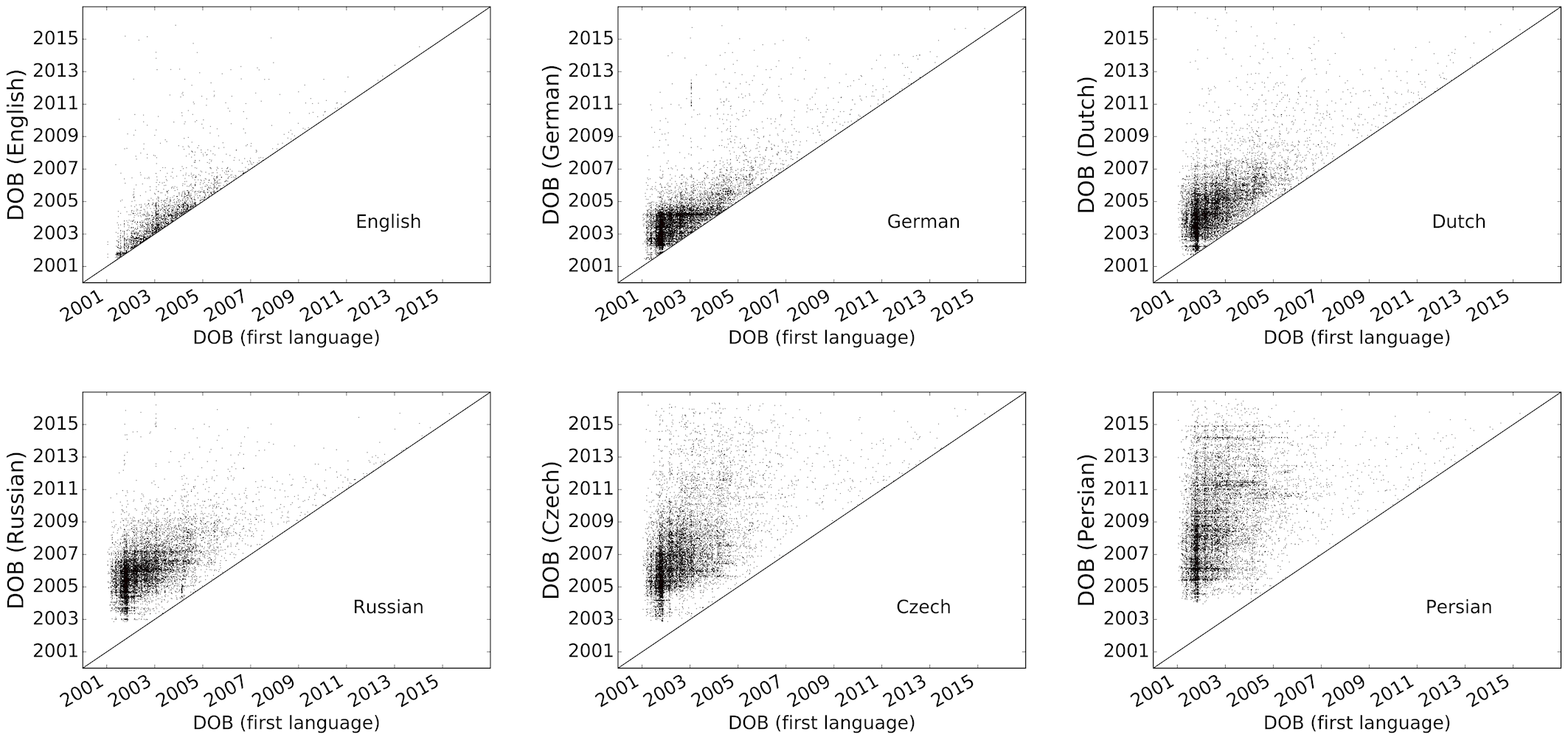}}
\caption{Scatter plots showing the original date of birth (in whichever language) of an article versus the date of birth in the language marked in each plot. Not surprisingly, English is the first Wikipedia for most of the articles, which is reflected in the majority of the dots being on or very close above the diagonal. Moving from the top left plot to the bottom right plot there is a clear trend of dots scattering further and further away from the diagonal, thus indicating that for these languages the bulk of the articles appear with a growing delay behind their first dates of birth (which is most often in English). The displayed six languages each belong to a different cluster that we discuss in the following figures and in the Table~\ref{tablelangclust}. English and German compose clusters of their own.}
\label{scatter}
\end{figure*}

Before turning to average growth patterns, we show in Fig.~\ref{scatter} scattered plots obtained for six different languages, where the DOB of an article in a particular language is displayed in dependence on its first DOB (in whichever language that is). Each article is represented by a small black dot. If a dot falls exactly on the diagonal, it means that this is the first language for that article. The vertical distance between a dot and the diagonal measures the difference between the DOB in the first language and the DOB in the language considered in each plot.

Looking at specific languages, it can be observed that most of the English articles fall either directly on or very close above the diagonal. This confirms that English is the language in which articles are most likely to appear first. As we move further to the right in the upper row, to the German and the Dutch, it can be observed that more and more dots fall significantly above the diagonal, thus indicating that for these two languages at least some of the articles appear with a considerable delay with respect to when they have appeared first. In the bottom row from left to right, for the Russian, Czech, and Persian, the same trend continues, to the point where barely any article falls on the diagonal, thus indicating that the Persian language was never the ``mother language'' of an article in our database. As we will show in what follows, the six displayed languages actually belong to six different language clusters, which share notable statistical similarities in their growth patterns.

\subsection{Clustering by cumulative number of articles over time}

\begin{figure}[b]
\centering{\includegraphics[width = 8.5cm]{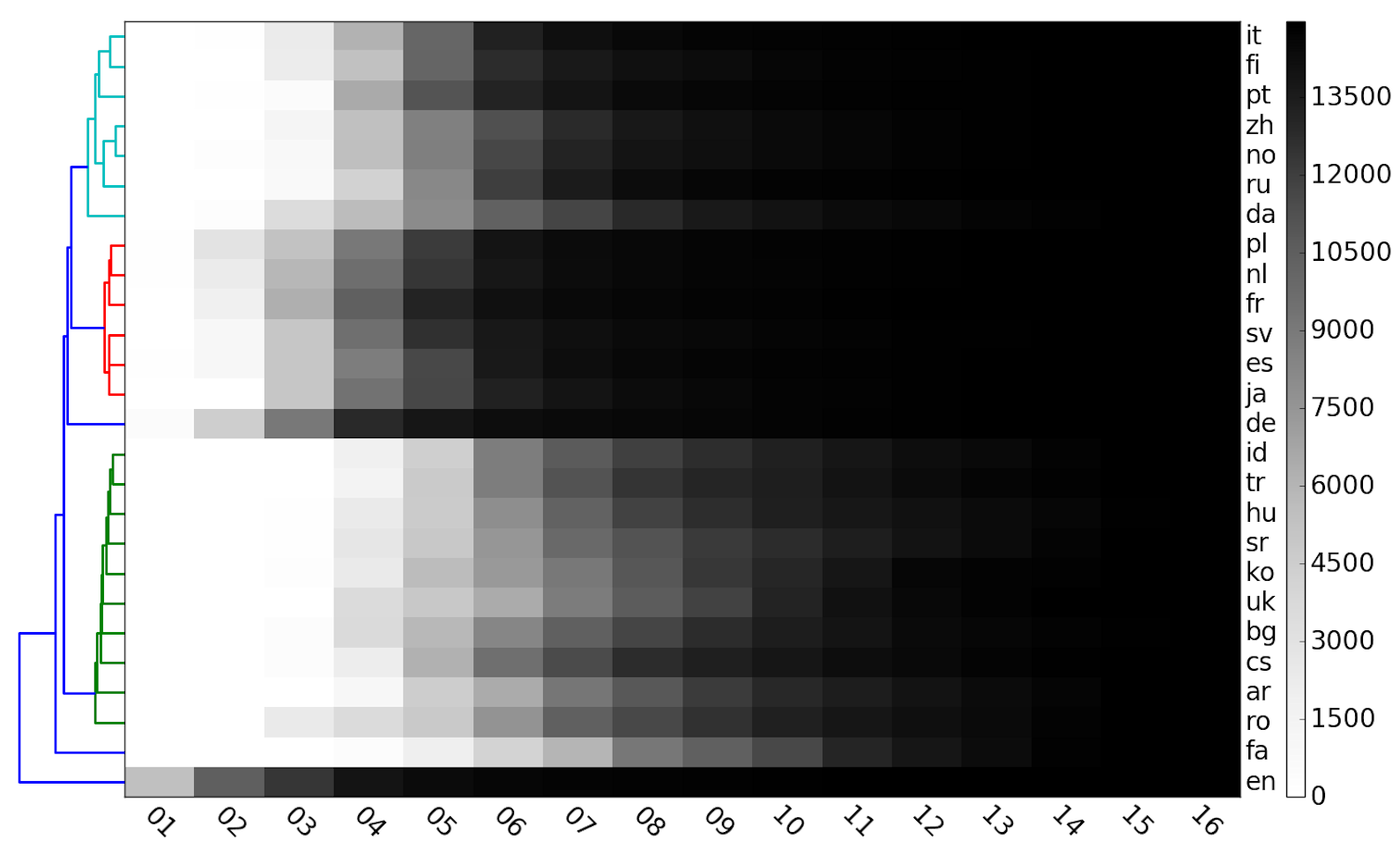}}
\caption{This analysis yields the clusters listed in the first (leftmost) column of Table~\ref{tablelangclust}. The gray-scale in each block encodes the cumulative number (see the bar on the right) of articles that appeared up to that year (horizontally) in that language (vertically). Considering the example of English language, it can be observed that most of the articles considered in our analysis were in existence by the year 2005. This is closely followed by the German articles, then by the French articles, and so on. Notably, this is valid only on average, while each individual article follows a random path as shown in Fig.~\ref{random}. The dendrogram on the left shows the clustering based on the displayed growth rates, such that the languages within each cluster grow similarly fast. Colors in dendrogram indicate the obtained clustering. For the abbreviation of languages see Table~\ref{table26lang}.}
\label{growth}
\end{figure}

To study regularities in the average growth patterns over different languages, we show in Fig.~\ref{growth} a gray-scale heatmap where the shade of each block encodes the cumulative number of articles that appeared up to a given year in a particular language, where years are displayed horizontally and languages vertically. Unlike the random trajectories displayed in Fig.~\ref{random}, here it can be observed that, on average at least, some Wikipedias grow much in the same way as others.

By determining the dendrogram that links together languages that exhibit similar growth rate~\cite{hastie_09}, we find that the 26 languages can be categorized into six clusters. Namely, looking at dendogram in Fig.~\ref{growth} (dark blue lines in particular), we note that classifying the growth rates into six groups (clusters) makes the clearest distinction among those groups. The vertical ordering of languages in Fig.~\ref{growth} comes out of the dendrogram, which is shown on the left margin of the heatmap. From this dendrogram we obtain the clusters that are reported in the first column of  the Table~\ref{tablelangclust}.

While it is perhaps expected that the English language would be in a cluster of its own due to its prominence and widespread use around the world, it is nevertheless surprising to observe that for those clusters that contain more than one language, the languages have very little in common in terms of the language family and group they belong too (see Table~\ref{table26lang}). The simplest explanation for the make-up of the clusters, which would be that language similarity and thus the ease of translating give rise to similar growth rates, does not apply. Other columns in the Table~\ref{tablelangclust} report clusters obtained via different methods, as we explain in the reminder of this Section.

\subsection{Clustering by average DOB difference}

\begin{figure}[b]
\centering{\includegraphics[width = 8.5cm]{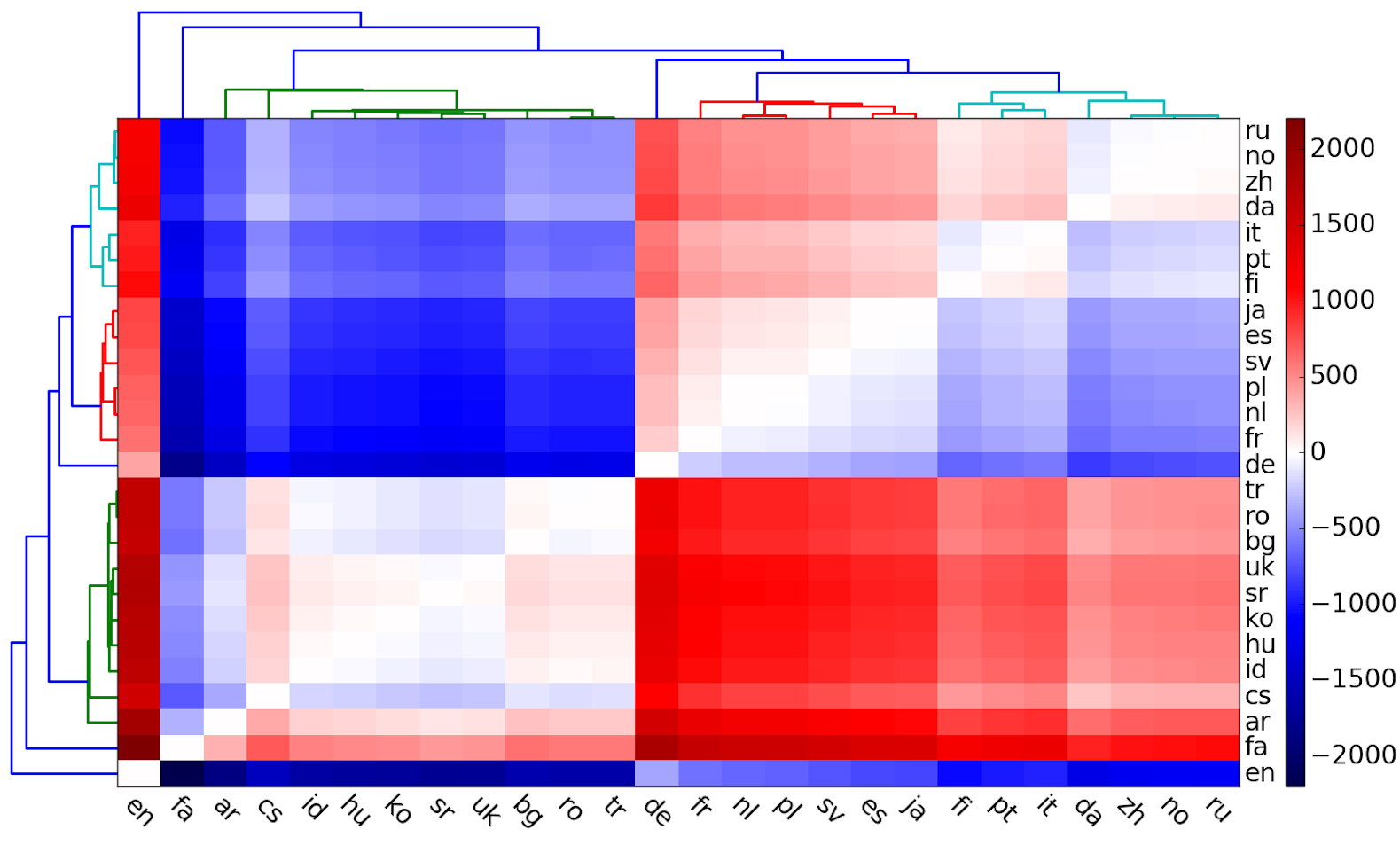}}
\caption{ This analysis yields the clusters listed in the second column of Table~\ref{tablelangclust}. The color of each block encodes the difference in days between the appearances of articles in two respective languages (see color bar), averaged over the entire ensemble of articles. The colors in the plot show which Wikipedias are running ahead and which are running behind in a pairwise comparison. White color indicates that the two languages are practically synchronized, i.e., that articles in those two Wikipedias on average appear simultaneously. It can be observed that the English language is running ahead of all the other languages, while the Persian language is running behind all of them. Also visible are several groups of languages that are well synchronized among them, composing clusters of languages as indicated in the dendrogram on the left and top (the two are identical). Dendrogram is again cut the have six clusters. For the abbreviation of languages see Table~\ref{table26lang}. Despite of the methodological differences, the 26 languages always fall into much the same clusters as observed before in Fig.~\ref{growth} (see Table~\ref{tablelangclust} for direct comparison).}
\label{birth}
\end{figure}

Of course, determining the clusters based on average growth rates is perhaps not the best, and certainly not the only way to find shared properties among Wikipedias. Arguments could thus be raised whether a different method of clustering would yield a more expected outcome in terms of grouping of Wikipedias. With this in mind we now examine a different clustering approach. For each pair of Wikipedias we considered the difference between DOBs in those two Wikipedias for all articles. Averaging these differences over the entire ensemble of articles, we obtain the heatmap shown in Fig.~\ref{birth}, where the color of each block indicates the average DOB difference between the respective pair of languages (Wikipedias). A certain language is on average either ahead (red) or behind (blue) another language, while (almost) white blocks denote that a given pair of languages is practically synchronized, i.e., that articles -- on average -- appear simultaneously in both Wikipedias.

Using these information as input we obtain another clustering whose dendrogram is shown on the margins of the heatmap in Fig.~\ref{birth}. These clusters are listed in the second column of Table~\ref{tablelangclust}. Although some differences exist in comparison to the clusters determined via previous approach from Fig.~\ref{growth}, their content is largely the same, and again lacks correlation with the language families and groups listed in Table~\ref{table26lang}. We may thus reiterate the earlier conclusion that the growth of Wikis across different languages is likely governed by a complex interplay of factors beyond languages themselves.

\subsection{Clustering from absolute article delays}

\begin{figure}[b]
\centering{\includegraphics[width = 8.5cm]{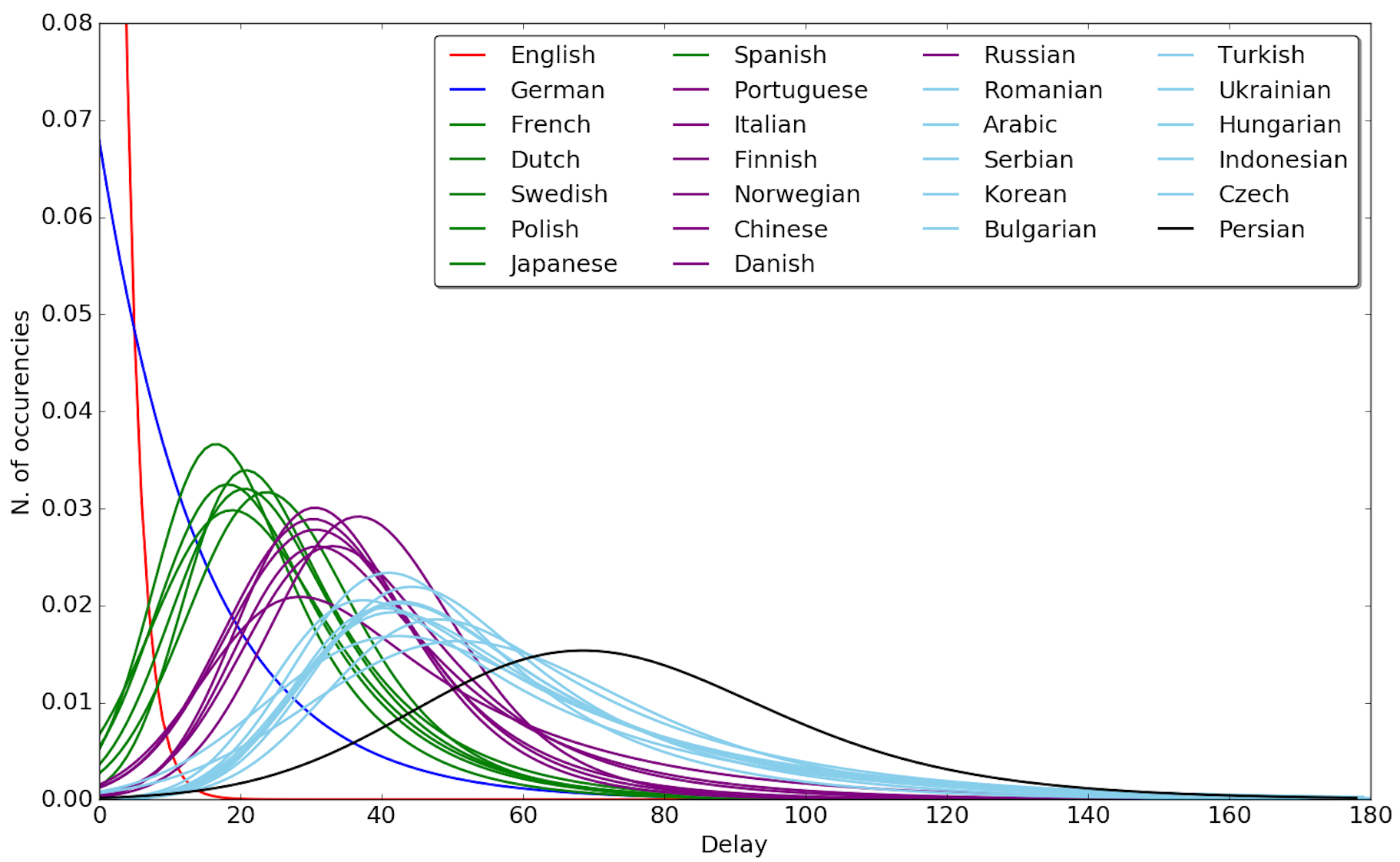}}
\caption{ This analysis yields the clusters listed in the third column of Table~\ref{tablelangclust}. Each curve corresponds to one Wikipedia, and it shows the distribution of delays (in days) of its articles behind the first appearance of these articles (in whichever language the first appearance happens to be). For clarity and simplicity, we do not show the actual data (due to being rather noisy), but we instead show the curves obtained by fitting the actual data to the exponentially modified Gaussian curves, since this way the clustering patterns are more clear. The fits were determined optimally by the shape of the actual data histograms. Different colors are used to distinguish different clusters of languages. Again, English and German are in their separate individual clusters, while the rest of the languages are clustered similarly to what we have already observed in our earlier analysis. The clustering results do not change if we use a different fitting function. This finding further confirms that the composition of the six language clusters is robust and largely independent of the clustering method.}
\label{delay}
\end{figure}
 
In contrast to averaging pair-wise Wikipedia DOB differences, next we examine the histograms (distributions) of absolute article delays for all Wikipedias. To that end we consider the difference between article's DOB in a specific language and that article's first DOB (in whichever language it happens to be). We thus obtain delay histograms, which for each Wikipedia capture the distribution of delays of its articles behind their first DOBs. These distributions are shown in Fig.~\ref{delay}, where each curve is an exponentially modified Gaussian curve, best fitted on the actual histograms (for simplicity and clarity of comparison we do not show the histograms but only the fitted curves). Clearly, the fact that English is the most common first language is reflected by its distribution being almost perfectly peaked at zero delay. This is true for German to a lesser extent, while the remaining Wikipedias can be neatly grouped into clusters, such that delay distributions within each cluster are remarkably similar in shape, size and peak value of delay. Note that clustering was in this case done using the distances between the curves as the similarity measure, in contrast to previsous cases.

Wikipedias in Fig.~\ref{delay} are marked by groups of different colors to reflect their organization into clusters that are obtained from this analysis, and are shown in the third column in Table~\ref{tablelangclust}. Again, we note striking similarities in the composition of clusters between this method and other two methods employed so far, despite these methods being fairly independent from one another.

\subsection{Clustering via multi-dimensional scaling}

\begin{figure}
\centering{\includegraphics[width = 8.5cm]{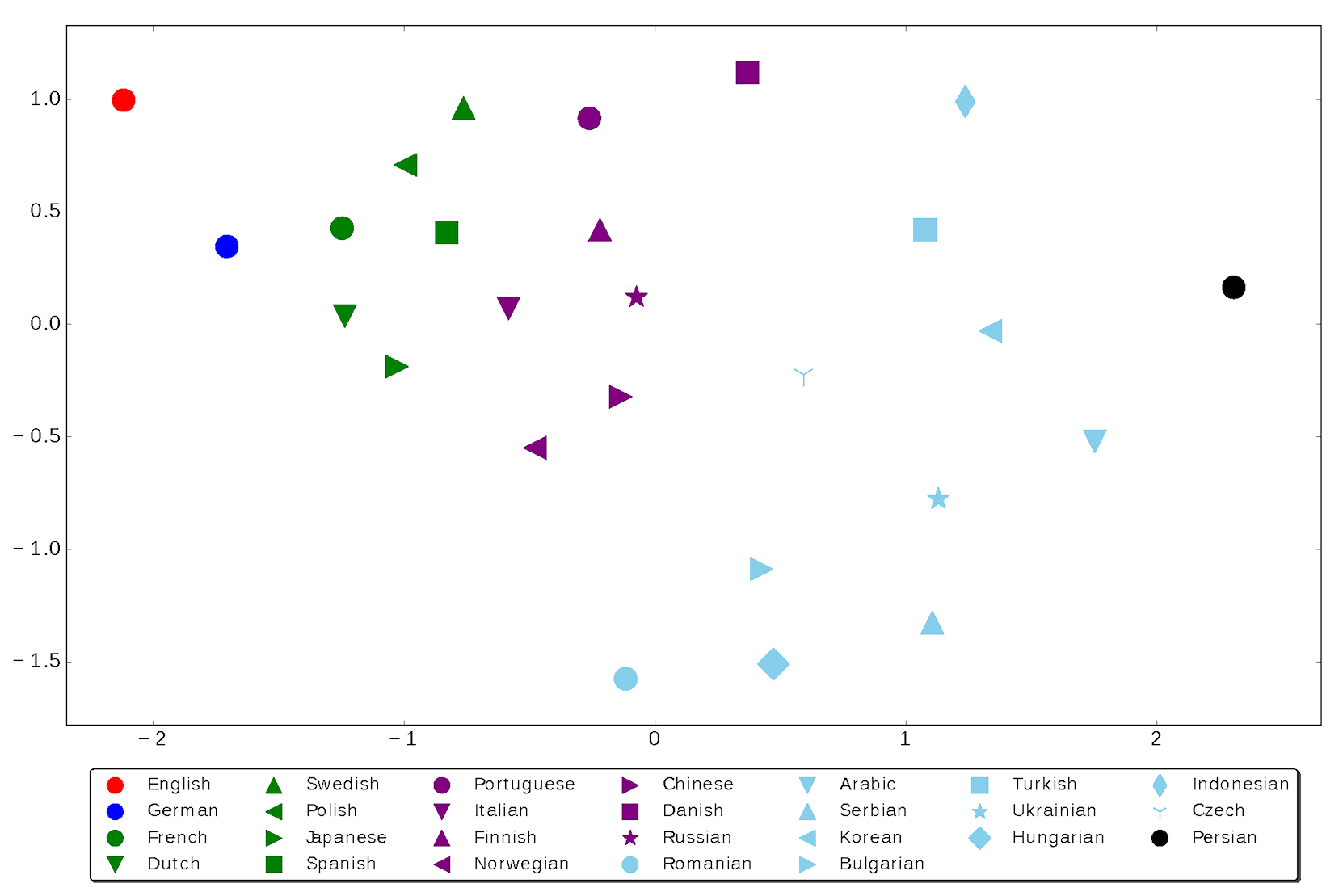}}
\caption{This analysis yields the clusters listed in the fourth column of Table~\ref{tablelangclust}. Each language (Wikipedia) is represented as a point in 2D space (see legend), while the geometric distance between any pair of points (Wikipedias) captures the averaged difference of DOBs of their articles (values used also in Fig.~\ref{birth}). Two Wikipedias close together are practically synchronized, while two Wikipedias far apart have one of them running ahead/behind the other. The MDS algorithm visualizes the points in 2D while keeping constant the ratios of their respective distances. Clusters of very similar composition can be identified, with English and German again separated from the rest. Each color indicates languages in one cluster. Note that because of the choice of visualization 2D space, points in some clusters (e.g. cyan) appear further from one another than they really are. This re-confirms the robustness of the obtained clusters to the method utilized for their determination.}
\label{mds}
\end{figure}

We conclude this Section by presenting yet another clustering approach, conceptually different from the previous ones. We go back to the data behind the Fig.~\ref{birth}, which consists of the average difference between DOBs (value measured as number of days) for each pair of Wikipedias. These values can be understood as pair-wise ``distances'' between Wikipedias, in the sense that two synchronized Wikipedias will be at distance zero from each other, while a pair Wikipedias that run ahead/behind each other will be at a certain positive distance from each other, which expresses to what extent are they not synchronized. Considering the Wikipedias endowed with pair-wise distances among them, we apply the standard multi-dimensional scaling algorithm (MDS)~\cite{levnajic2014} and represent the languages as points embedded in 2D space, as shown in Fig.~\ref{mds}. MDS transforms the set of pair-wise distances keeping the ratio between each two distances into a new set of pair-wise distances that can be embedded into a space of given dimensionality. By this procedure we are able to visualize the 26 Wikipedias as 26 points in space with distances between them illustrating the time delays between each pair of Wikipedias.

We can clearly see the same languages, where English and German are well separated from the rest, each defining its own individual cluster. Remaining languages can be grouped (this time using geometric considerations) into well localized clusters, each indicated by one color in Fig.~\ref{mds}. The clustering coming from this analysis is reported as last (fourth and rightmost) column in Table~\ref{tablelangclust}. And yet again, the composition of all six clusters does not substantially differ from the clusterings so far observed, despite a very different approach used this time.

\section{Discussion}

\begin{figure}[b]
\centering{\includegraphics[width = 8.5cm]{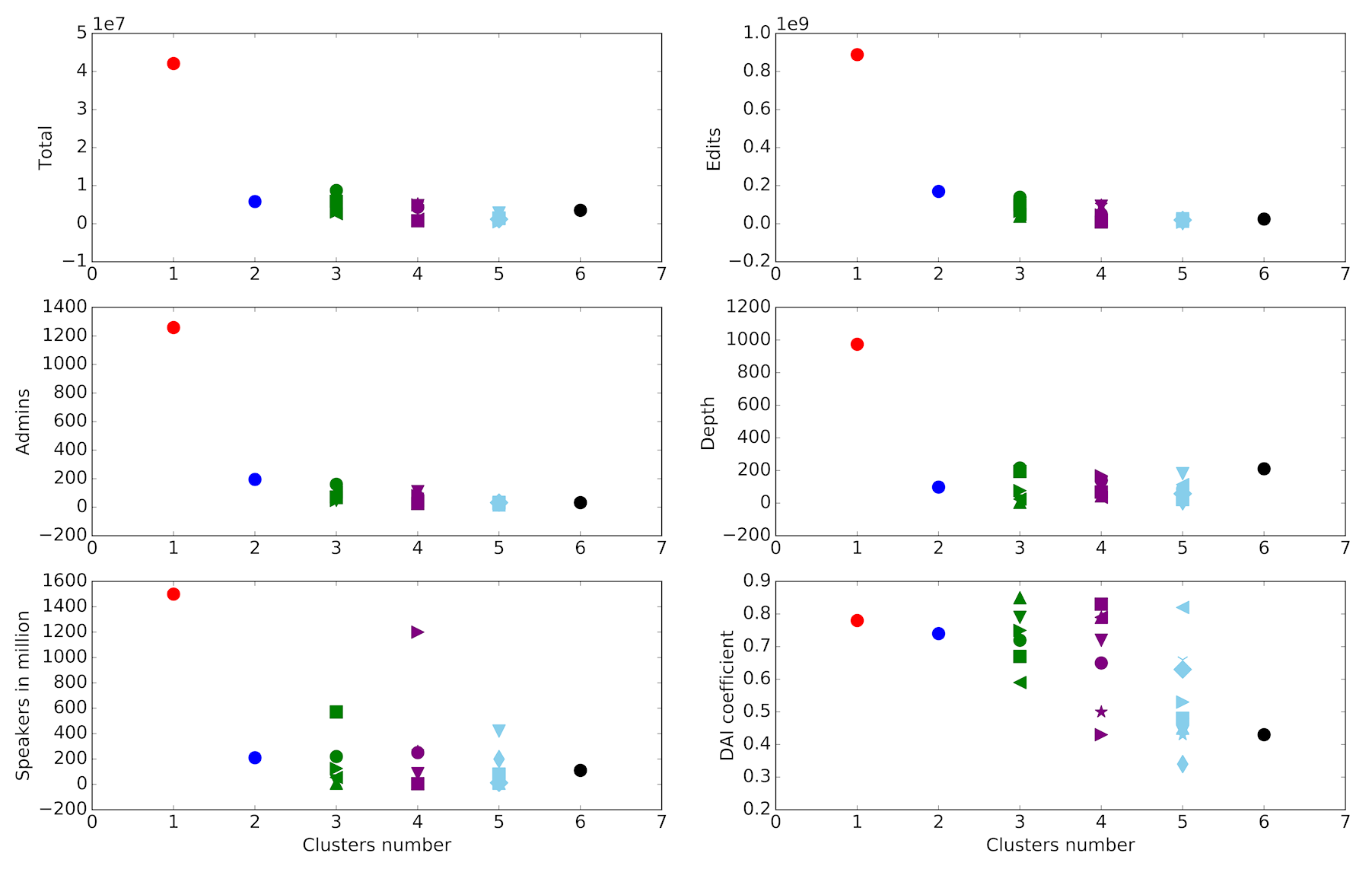}}
\caption{To examine whether the emergence of clusters can be explained by other independent data, we plot six scatter plots where for each language the cluster number (simply from 1 to 6) is on the horizontal axis, while the value for one of the six considered variables is on the vertical axis. The clustering scheme is as in Fig.~\ref{mds}. Six variables we consider are (from top left to bottom right): Total: the total number of Wikipedia pages including both articles and non-articles (e.g. images, talk pages etc); Edits: the number of edits (modifications) in a Wikipedia; Admins: the number of administrators of a Wikipedia; Depth: a proxy for the content quality of a Wikipedia defined as (Edits/Articles) $\times$ (Non-Articles/Articles) $\times$ (1 - Stub-ratio); Speakers in million: the total number of active speakers of a language; DAI coefficient: Digital Access Index (www.itu.int/ITU-D/ict/dai), a proxy for the IT literacy and accessibility to IT services of a population. All values are obtained independently from the main Wikipedia data. Having the clusters (colors) separated clearly along the vertical axis would amount to a good correlation. However, in the top two and the middle two plots (Wikipedia editing parameters) we see a weak correlation that decays towards larger clusters. In the bottom two plots (social parameters) we basically find no correlation. This analysis indicates that while some of these variables indeed explain the cluster structure to a modest extent, the complete and conclusive explanation is likely to be hidden in a diverse range of social, economic, IT related, and other factors.}
\label{interpretation}
\end{figure}

We have used a relatively small subset of Wikipedia articles that jointly exist in 26 different languages in order to study the commonalities of statistical growth patterns that are shared by Wikipedia editions in different languages. An individual article in general follows a random path over time as it is ``translated'' from one language to another (we put ``translated'' in quotes as we do not maintain that Wikipedias exclusively grow by translating from one another, but also by writing genuinely new articles). However, upon averaging over the dataset containing 14962 articles, statistical similarities emerge that quantify how Wikipedias grew in different languages over the last 15 years. In particular, we have observed robust clustering of languages into six distinct clusters, the composition of which is largely independent of the method used for the clustering analysis. This suggests that the Wikipedias that share a cluster indeed also share the same overall growth properties, which raises the question as for the reasons behind this. We also verified that considering one or two languages more or less does not have a significant impact on the cluster structure (six clusters can still be identified, but their content depends on which languages we remove from these 26, or which new ones we add on top of these 26).

The simplest explanation revolves around common linguistic or cultural traits that might account for similar ``translatability'' of articles. But unexpectedly, our research reveals that for those clusters that contain more than one language, these languages and countries/cultures behind them have very little in common, almost as if the clusters were formed randomly. It is indeed very hard to find any linguistic similarities between Indonesian, Turkish, Hungarian, Korean, Ukrainian, Czech, Arabic and Romanian language, which are placed in the same cluster by all examined clustering methods. The same is true if we look for possible common cultural traits between populations speaking these languages. This invites the conclusion that the growth of Wikipedia across different languages is governed by a series of factors other than linguistic or cultural familiarity. We arrive at similar conclusion looking at other multi-language clusters, which is again in favor of a strongly faceted array of factors that together contribute to the formation of the observed clusters.

To test this conclusion more quantitatively we obtained independent data on six variables that could account for emergence of clusters. Those are: the total number of Wiki pages, the number of edits in a Wiki, the number of administrators of a Wiki, ``depth''of a Wiki (proxy for the content quality), the number of active speakers of a language, and the Digital Access Index coefficient (proxy for the IT literacy/accessibility). As we show in Fig.~\ref{interpretation}, none of these variables offers a clear and conclusive explanation of the observed clustering structure, although some variables (e.g. the first three) do show a weak correlation with the clustering structure.

We therefore hypothesize that the observed similarities stems from a complex interplay of several social, economic and political factors, probably including at least the following three. First, the total population that regularly communicates in a given language, which roughly overlaps with the joint population of the countries where a given language is in use (while some of the considered languages are spoken in a single country, languages like English and German are in regular use in several countries). Second, the access to the Internet and average Internet literacy in a country, which has to do with the country's policy towards IT education, which in turn correlates with country's level of technological and economic development or even its political order. Thirdly, the general attitude towards importance of knowledge and education, from which comes the willingness and motivation to volunteer as Wikipedia editor in one's own language, and which in turn may depend on how strong are the emotions about ``national culture'' in a given country. Since estimating any of these factors is very difficult and the data is not immediately available, we at present cannot offer a conclusive proof of this hypothesis, which we leave as an interesting open question for future work. However, even if we recognize that there are likely more relevant factors behind this process, we note our relatively unexpected findings indicate that the complete set of influencing factors might actually be inferable.

Looking ahead and taking into account very interesting recent developments in research on multilingual Wikipedia \cite{eomplos15, yasseri13, yu2014pantheon, iniguez2014modeling, kim2016understanding, samoilenko2016linguistic, yun2016early}, we are certain that the time is ripe for data-based research on similar databases. This research should be primarily focused on the differences and synergies that emerge as a results of the interactions between different cultures in a world-wide collaborative environment and rely on a large volume of literature on collective social processes~\cite{wang2016statistical, matjaz2016} and spontaneous phenomena such as crowdsourcing~\cite{lee2013, guazzini2015}. Except data-based approaches and statistical modelling, newer IT developments allow for social experiments with several hundreds of participants to be carried out under controlled conditions~\cite{jelena2010, jelena2012}. Our vision is that integrated methodological framework based on data-driven models on one side and controlled social experiments on other side can lead to an interdisciplinary platform for systematic study of collective social phenomena. Constructive synergy between social and computational sciences should here play the crucial role, since the applicative front of social processes is virtually infinite~\cite{lazer_s09, jelena2010, lee2013}.

\begin{acknowledgments}
This research was supported by the Slovenian Research Agency via projects J1-7009 and V5-1657, programs P1-0383 and P5-0027, the Young researcher scheme 36664, and by the EU via MSC-ITN-EJD consortium COSMOS 642563.
\end{acknowledgments}

\end{document}